# Control of Electrons' Spin Eliminates Hydrogen Peroxide Formation During Water Splitting

Wilbert Mtangi[1], Francesco Tassinari[1], Kiran Vankayala[1], Andreas Vargas Jentzsch[2], Beatrice Adelizzi[2], Anja R.A. Palmans[2], Claudio Fontanesi[3], E.W. Meijer*[2], and Ron Naaman*[1]

[1]Department of Chemical Physics, Weizmann Institute of Science, Rehovot 76100, Israel;
[2]Institute for Complex Molecular Systems, Eindhoven University of Technology, Eindhoven, the Netherlands;
[3]Department of Engineering, University of Modena and Reggio Emilia, Via Vivarelli 10, 41125 Modena, Italy

**ABSTRACT:** The production of hydrogen through water splitting in a photoelectrochemical cell suffers from an overpotential that limits the efficiencies. In addition, hydrogen-peroxide formation is identified as a competing process affecting the oxidative stability of photoelectrodes. We impose spin-selectivity by coating the anode with chiral organic semiconductors from helically-aggregated dyes as sensitizers; Zn-porphyrins and triarylamines. Hydrogen peroxide formation is dramatically suppressed, while the overall current through the cell, correlating with the water splitting process, is enhanced. Evidence for a strong spin-selection in the chiral semiconductors is presented by magnetic conducting (mc-)AFM measurements, where chiral and achiral Zn-porphyrins are compared. These findings contribute to our understanding of the underlying mechanism of spin selectivity in multiple electron-transfer reactions and pave the way towards better chiral dye-sensitized photoelectrochemical cells.

## ■ INTRODUCTION

Since it has no carbon, has the highest specific enthalpy of combustion of any chemical fuel, and generates water as its oxidation product, hydrogen has been referred to as the fuel of the future.[1] Although significant progress has been made over the past decades,[2] the generation of hydrogen by green, sustainable methods on a global scale remains in the future.[3-8] While technologies exist for the electrolysis of water and photo-electrochemical generation of hydrogen from water,[9] the processes involve significant overpotentials and the formation of peroxides and superoxide radical by-products. These by-products have the tendency to adsorb onto the photo-catalyst, poisoning it, thereby reducing its stability and lifetime.[10] Although specific catalysts are proposed to use the peroxides as intermediates,[11] this path requires higher voltage. Therefore, a fundamental solution for the off-pathway products is essential.

Water splitting is a four-electron process that generates hydrogen molecules having singlet ground states and oxygen molecules having triplet ground states. Commonly, the artificial water splitting process requires an overpotential of about 0.6 V vs. normal hydrogen electrode (NHE), to drive the oxygen evolution reaction.[12-14] The importance of electron-spin correlation of electrons in generating $O_2$ has been debated for biological photosynthesis. Particularly the chemistry associated with the oxygen evolution reaction in photosystem II has been examined.[15-17] Nevertheless, the details of the mechanism of the O=O bond formation remain unresolved.[18-20] Although the spin state of the electrons involved is rarely discussed in works exploring artificial photosynthesis, recent theoretical studies suggest that the overpotential required to split water is linked to restrictions on the electrons' spin in generating a ground state triplet oxygen molecule.[21,22] In recent experimental work, it has been shown that when the anode in the water splitting cell is coated with chiral molecules, the overpotential is reduced.[23] It has been proposed that the effect is due to spin filtering occurring when electrons are conducted through chiral systems.[24] However and very importantly, the possible role of the spin control in suppressing the formation of hydrogen peroxide has not been discussed nor experimentally addressed.

Here, we hypothesize that hydrogen peroxide is produced due to uncontrolled spin alignment and greatly contributes to the high overpotentials. Hence, controlling the spin state of the electronic potential on which the reaction occurs should result in more efficient oxygen production and limited production of hydrogen peroxide. Although, we do not strive here to present the highest production of hydrogen and oxygen – optimizing the cells is ongoing – the results presented show an unprecedented control of chemical kinetics through spin selection.

## ■ RESULTS AND DISCUSSION

To first control the spin state, the ITO anode in the photoelectrochemical cell was coated with two families of organic semiconductors, Zn porphyrins and tri(pyrid-2-yl)amine trisamide, TPyA, both in their chiral (using enantiomerically pure side chains) and achiral (using achiral side chains) versions (Fig. 1A,B). By modifying the side chains of the molecules, we control the resulting helical supramolecular assemblies into either a bias for one helical sense or a racemic mixture of both helical senses, respectively. Thus, in the present study, electrons are transmitted into the $TiO_2$ substrate through the same molecular system, which only differs in its molecular organization; either one helical sense or a mixture of both. In the case of Zn-porphyrins, the 390 nm Soret band confirms the formation of helical supramolecular structures in solution (Fig. 1C),[25,26] which are retained when transferred to the surface (inset of Fig. 1C). In a similar manner, the 317 nm band is indicative of the formation of a supramolecular assembly for the TPyA molecules. For both molecules, the chiral analogues show a strong CD response which is not observed in the achiral version (Fig. 1E, F). The achiral molecules most probably aggregate in equal amounts of left- and right-handed helices (racemic mixture), while the chiral molecules prefer one of both only.

**Insert Figure 1 here**

Magnetic conducting atomic force microscopy (mc-AFM) measurements were conducted to verify the spin selectivity of electron transmission through the Zn-porphyrin stacks.[27] Fig. 2B shows the dependence of the current on the orientation of the magnetic field at the tip of the mc-AFM for the stack of chiral porphyrins, which confirms that one spin is preferred over the other. It is important to realize that the ratio between the two spin currents is affected by the non-ideal spin injection from the magnetic tip. Since the spin polarization of the tip is only about 35%, it means that the actual spin filtering of the molecular system is about 4:1, in other words only about 20% of the electrons conducted through the chiral molecular aggregates in this experiment have the "wrong" spin. From Fig. 2C, it can be observed that for the achiral compounds, the magnitude of the current measured as a function of applied bias does not depend on the orientation of the magnetic field; hence the conduction through this system with equal amount of right and left handed helices is not spin specific. These results indicate that the supramolecular structures formed from the chiral porphyrins can efficiently filter spins and are consistent with previous reports in which chiral molecules have been observed to be good spin filters.[28]

**Insert Figure 2 here**

Next, photoelectrochemical measurements were conducted in a three-electrode cell, with the Ag/AgCl (saturated KCl) as the reference electrode and a Pt wire as the cathode (Fig. 3). A 0.1 M $Na_2SO_4$ (pH = 6.56) aqueous solution was used as the electrolyte. $TiO_2$ substrates fabricated and functionalized as outlined in the supplementary material, were used as photo-anodes. In these cells, the magnitude of the measured current is correlated with the amount of oxygen bubbles produced at the anode and hydrogen bubbles at the cathode.[23]

Higher photocurrent densities are observed for photoelectrodes functionalized with helical aggregates of chiral molecules with preferred helicity, compared to those coated with racemic aggregates of achiral ones. This is remarkable, given that the chemical compositions of the chiral and achiral molecules for the two sets used in this study are very similar, except for the stereo-center present in the chiral molecules. In general, the magnitude of the photocurrents obtained with the Zn-porphyrins are typically low, as has also been observed by Moore *et al*.[29] With the TPyA molecules, the photocurrent densities are reasonably high, since UV light is used for illumination of the cell. The activity of the photoelectrodes is known to depend strongly on their electronic properties, therefore Mott Schottky measurements were performed to characterize the electronic properties of $TiO_2$ electrodes modified with aggregates of Zn porphyrin and TPyA molecules. A flat-band potential, $V_{bi}$ of -0.51V vs. Ag/AgCl was obtained in the dark for both the chiral and achiral Zn-porphyrins (Fig. 3E), while a value of approximately -0.70V vs. Ag/AgCl was obtained for the TPyA molecules (Fig. 3F), an indication that the difference in the photocurrents, for chiral and achiral species, is not related to the modification of the electronic properties of the photoelectrodes. Thus, the differences in the photocurrents must be attributed to the chirality of the molecules.

**Insert Figure 3 here**

After showing the difference in water splitting for the chiral versus racemic aggregates, the beneficial effect of electrons' spin control is highlighted by its effect on hydrogen peroxide formation. An indirect quantification of the $H_2O_2$ produced during photoelectrochemical water splitting was conducted. Spectrophotometric titration of the used electrolytes were performed using o-tolidine as redox indicator.[30,31] The amount of peroxide formed has been quantified through Ellms-Hauser method calibrating the system with commercial $H_2O_2$ (see S.I.). In the presence of $H_2O_2$, a yellow color appears with an absorption peak at about 436 nm. This peak is characteristic for the complete two-electron oxidation product of o-tolidine formed by the reaction with hydrogen peroxide.[32]

The electrolyte obtained from the bare $TiO_2$ and electrodes functionalized with achiral dyes showed the characteristic peak at 436 nm, indicating the production of $H_2O_2$ during water oxidation (see *table S1*). Fascinatingly, no detectable amount of hydrogen peroxide was observed for electrodes with the chiral molecules physisorbed (Fig. 4A, B). Noticeably, after 40 mins of irradiation, $43 \pm 5$ mmol$L^{-1}$ of hydrogen peroxide have been produced with the achiral Zn-porphyrin functionalized system while non-detectable levels of peroxide have been found using the chiral analogue. The TPyA functionalized systems show low absorbance intensity at 436 nm. The relatively low level of $H_2O_2$ in the case of A-TpyA may result for performing the electrochemistry measurements under illumination with UV light, which might have led to the disproportionation of the produced peroxide before titration. However, also in this case, much less $H_2O_2$ is produced with the chiral molecules (S-TPyA) then with the achiral ones (A-TPyA). Additional evidence is obtained by using achiral 3-mercaptopropionic acid and the chiral oligopeptide [(COOH)-(Ala-Aib)7-NH-(CH2)2-SH] (*fig. S23*). Only the latter decreases the $H_2O_2$ production. These results further indicate that the observed elimination of the hydrogen peroxide production is general for all chiral molecules.

**Insert Figure 4 here**

The results presented here in the quantification of $H_2O_2$ production together with the electrochemistry data show a strong correlation between the overpotential, the formation of $H_2O_2$, and the electron's spin alignment control. During water splitting, two $OH^-$ species must combine to form molecular oxygen in its triplet ground state. In the process, an electron from each $OH^-$ is transferred to the anode. This leaves the two $OH^\bullet$ radicals in their doublet ground state, namely each $OH^\bullet$ has one unpaired electron. When there is no spin control and the interaction electronic potential has a singlet character, the formation of $H_2O_2$ is possible (Fig. 4). However, when the electron's spins are aligned in a parallel fashion, the two electrons interact on the triplet potential surface which correlates with the formation of the ground state molecular oxygen and on which the formation of $H_2O_2$ is symmetry forbidden.

The formation of oxygen and hydrogen peroxide are anti-correlated; i.e., the formation of one is coming at the expense of the other. This is what the present results indicate. Of course, the substrate (the catalyst) may induce spin-orbit coupling that will mix the spin states of the OH radical. However, for $TiO_2$ and other relatively light materials, the spins of physisorbed molecules are expected to be conserved, as observed by XPS studies.[33] Note that substrate-induced spin-orbit coupling may explain the low overpotential observed for some oxides,[34,35,36] magnetic catalysts[37,38,39,40] or catalysts made from heavy atoms. For these systems, the side products are not significantly suppressed. Indeed, no investigations were performed in the past aimed at controlling the formation of $H_2O_2$.

## ■ CONCLUSIONS

The experimental results observed in the current study provide new insights into the mechanism behind oxygen formation in the oxygen evolution reaction and provide a new path for improving the efficiency of the water splitting process. Obviously, much work has to be done to optimize the system with other and more effective chiral dyes, chiral semi-conductors, and chiral catalysts, but a proof-of-principle to make another counter-intuitive approach is presented. The control of electrons' spin in the chemical kinetics during the photochemical water splitting will also rejuvenate the field of magnetic field effects in chemical kinetics and related phenomenon as reviewed by Steiner and Ulrich many years ago, following the discovery and understanding of electron spin polarized phenomena during chemical reactions[41]. This is the more of importance due to the recent interest in photoredox catalysis in organic chemistry with exciting enatioselctivities[42]. In addition, the work points to the importance of chirality and spin selectivity in multiple electron reactions in biology.

## ■ ASSOCIATED CONTENT

**Supplementary information**
This available free of charge

## ■ AUTHOR INFORMATION

**Corresponding authors**
ron.naaman@weizmann.ac.il
e.w.meijer@tue.nl
**Notes**
The authors declare no competing financial interest

## ■ ACKNOWLEDGEMENTS

RN acknowledges discussion with Profs. Andrew Rapp and David Waldeck and acknowledges partial support from the European Research Council under the European Union's Seventh Framework Program (FP7/2007-2013) / ERC grant agreement n° [338720]. The Eindhoven laboratory acknowledges financial support of EURO-SEQUENCES (H2020-MSCA-ITN-2014, No 642083) and the FMS Gravitation Program (No 024.001.035).

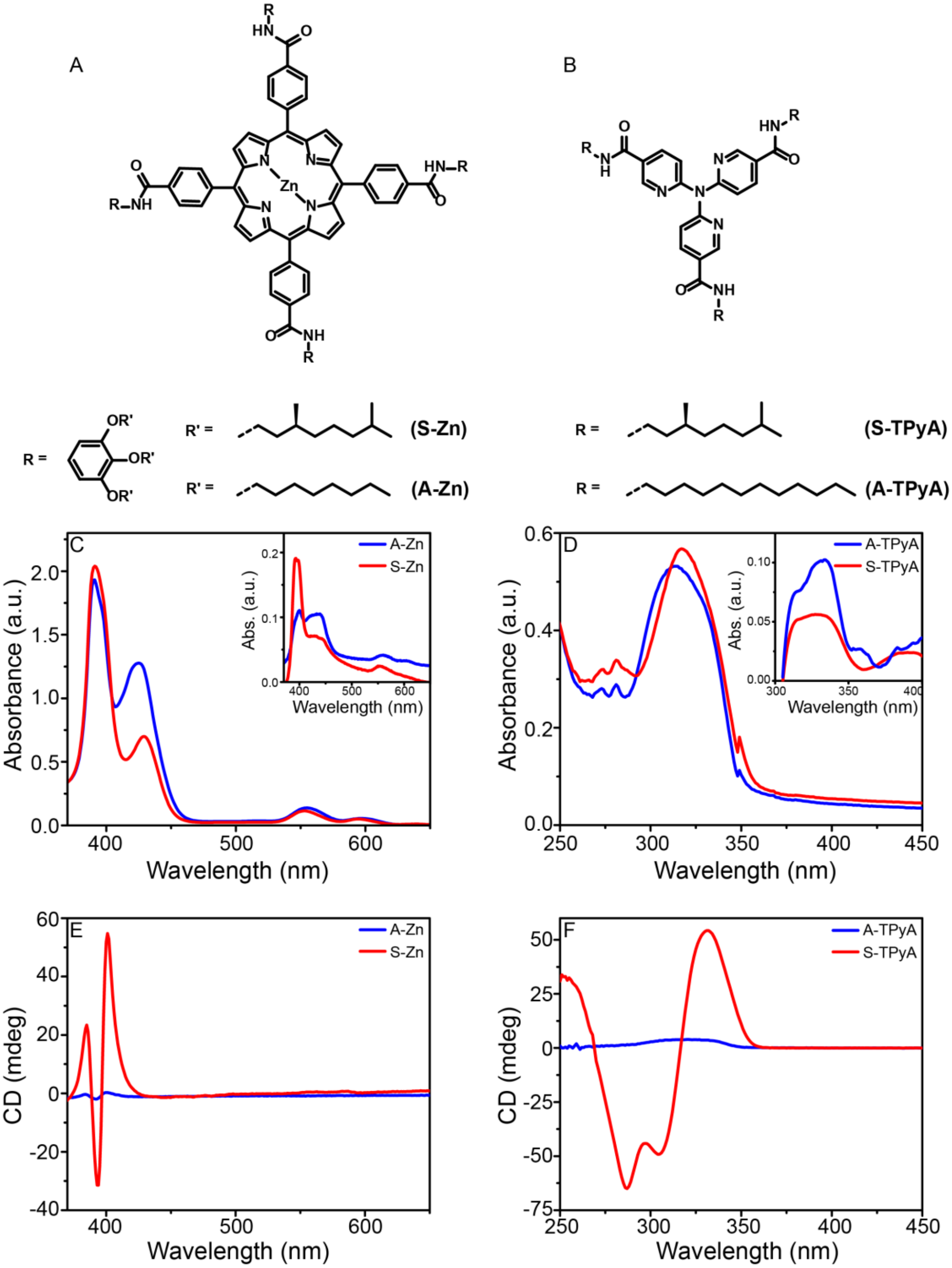

**Fig. 1 Molecules used as photosensitizer and their supramolecular aggregation**. (**A**) The chiral and achiral Zn porphyrins, **(C** and **E)** their absorption spectra and CD spectra of the aggregated state in solution (1.7x10$^{-5}$M in methylcyclohexane) and when adsorbed on the surface (inset) (**B**) the chiral and achiral TPyA, (**D** and **F**) their absorption and CD spectra of the aggregated state in solution and when adsorbed on the surface (inset). The red curves represent the chiral molecules, while the blue represent the achiral ones.

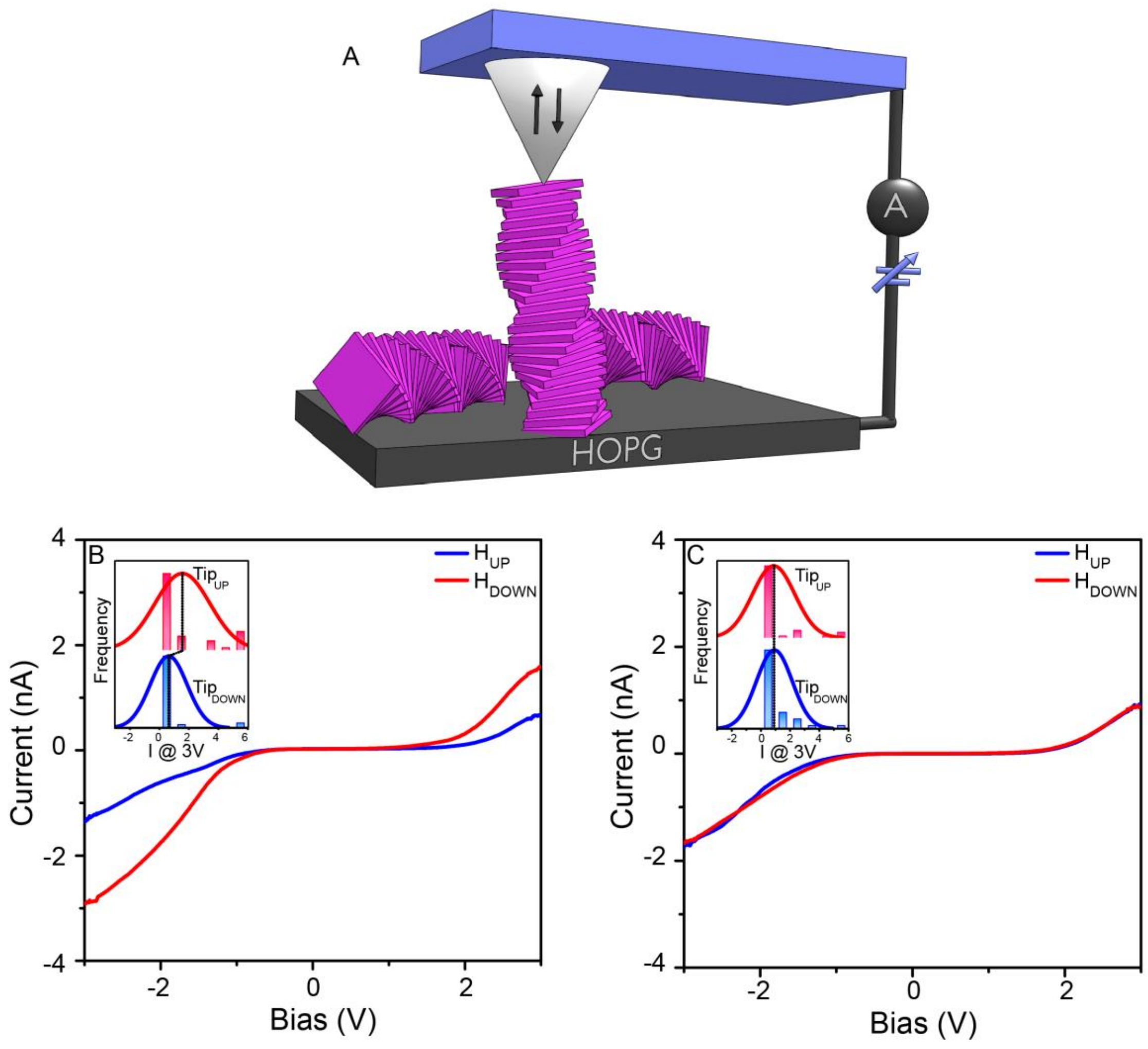

**Fig. 2** Magnetic conducting atomic force microscopy **(mc-AFM) measurements on stacks of chiral and achiral Zn-porphyrins.** (**A**) The experimental setup. The current as a function of the applied voltage obtained from the chiral (**B**) and achiral (**C**) Zn-porphyrin molecules. The insets show corresponding histograms of currents obtained at 3 V for the two opposite magnetic orientations of the tip.

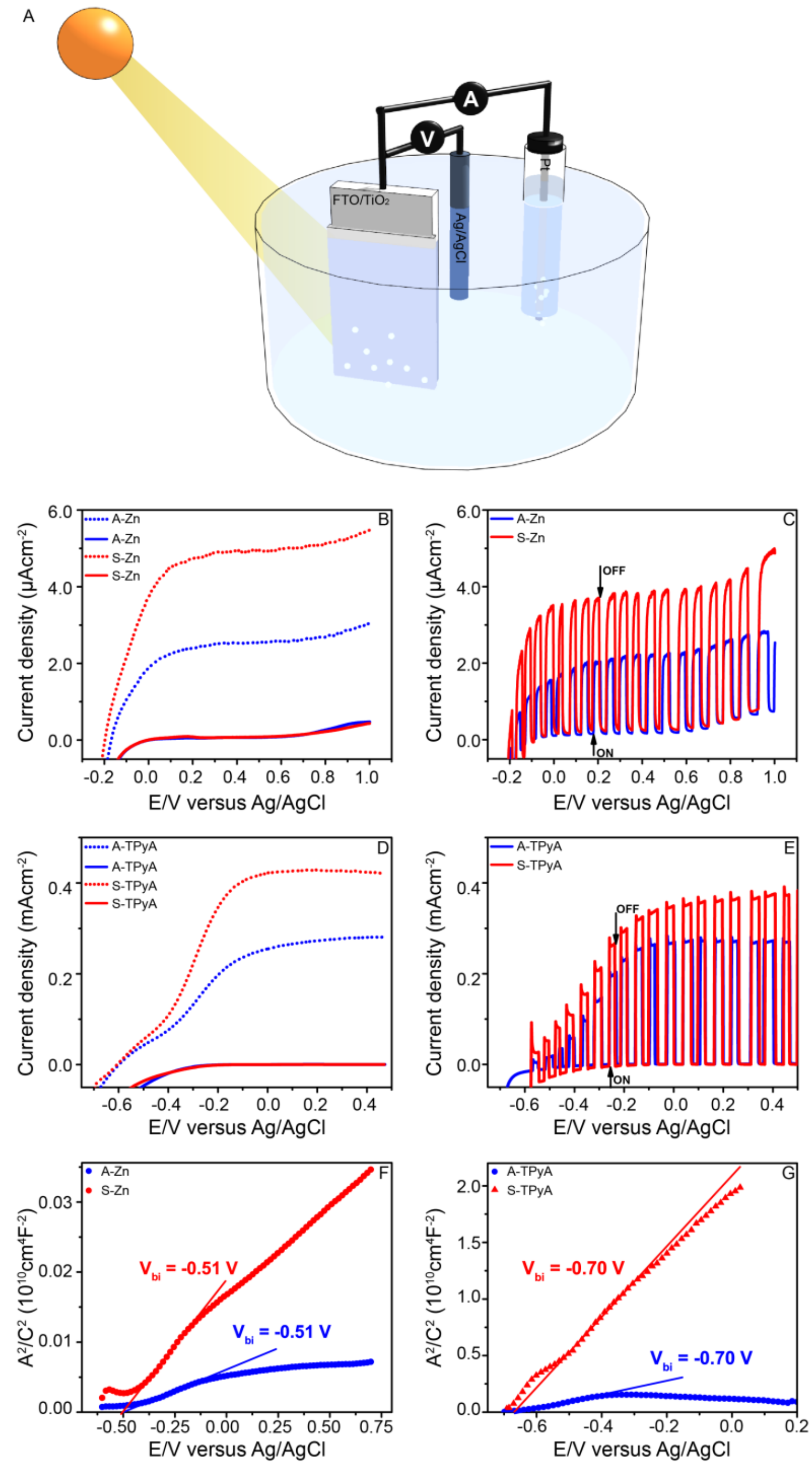

**Fig. 3 The photoelectrochemical cell and the current density as a function of the potential.** The potential is given vs. the Ag/AgCl electrode, when the $TiO_2$ electrode is coated with self-assembled achiral (blue lines) or chiral (red lines) molecules. In (**A**) and **(B**), Zn-porphyrins were used while in (**C**) and (**D**), TPyA molecules were employed. The measurements were performed in the dark (solid lines) and under illumination (dotted lines). In (**B**) and (**D**) measurements were performed while chopping the light. All measurements were performed at a scan rate of 10 $mVs^{-1}$. The flat-band potentials in the dark for Zn-porphyrin and TPyA molecules were obtained from Mott-Schottky plots (**E** and **F**, respectively) at a frequency of 1.99 kHz and oscillation voltage of 20 mV.

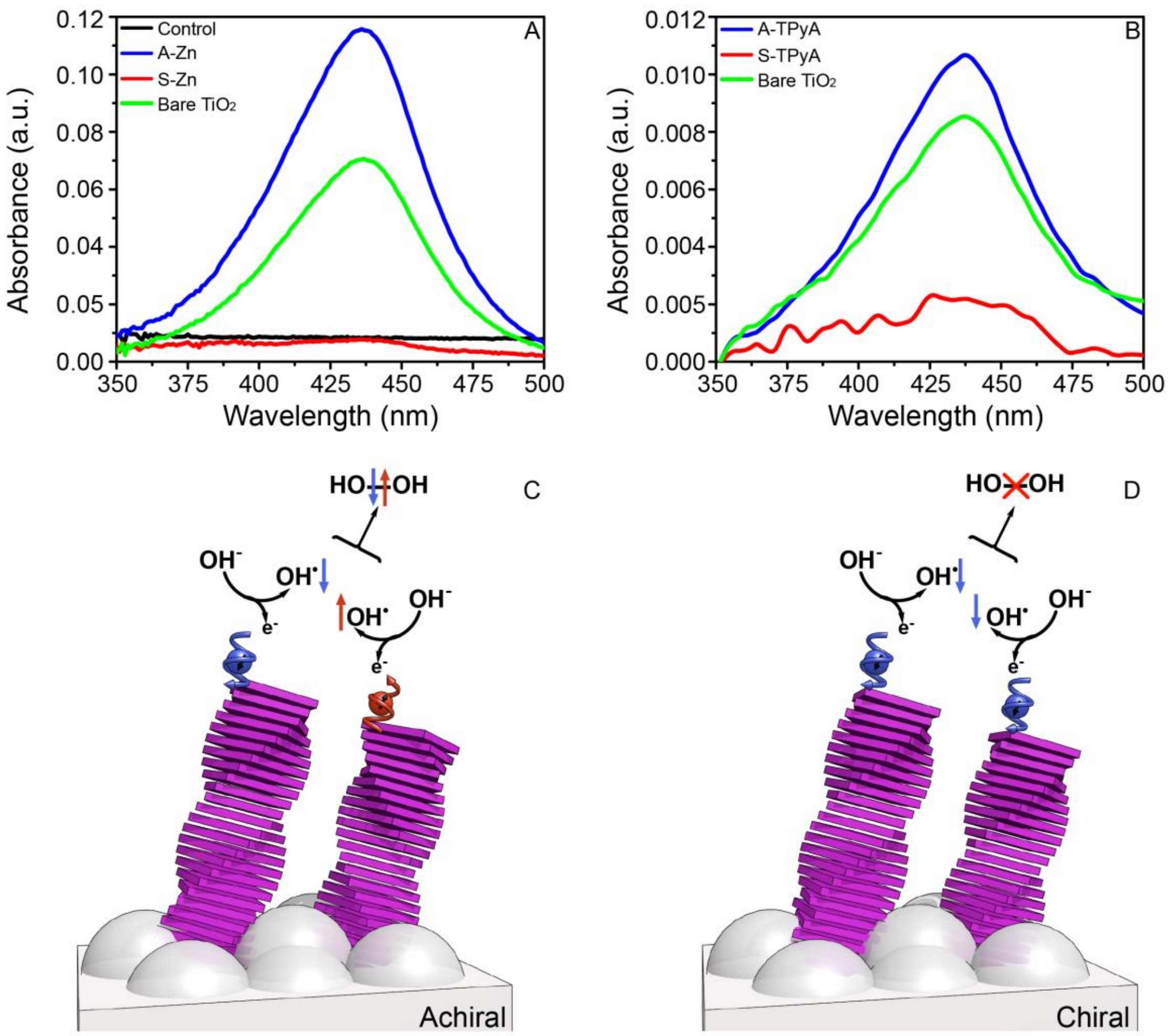

**Fig. 4 Control of the hydrogen peroxide production.** UV-vis absorption spectra from the titration of the used electrolyte ($Na_2SO_4$) with o-tolidine of bare $TiO_2$ and $TiO_2$ electrodes coated with (**A**) self-assembled Zn-porphyrins of either achiral (A-Zn) or chiral (S-Zn) and (**B**) TPyA molecules. The control refers to the titration of unused $Na_2SO_4$ with o-tolidine. (**C**) When the electrons transfer to the anodes is *non spin specific* the spins of the unpaired electrons on the two $OH^{\bullet}$ are aligned antiparallel, hence the interaction between the two $OH^{\bullet}$ is on a singlet surface that correlates with the production of hydrogen peroxide ($H_2O_2$). (**D**) When the electron transfer to the anode is *spin specific*, the spins of the two electrons are aligned parallel to each other, hence the two $OH^{\bullet}$ interact on a triplet surface that forbids the formation of $H_2O_2$ and facilitates the production of oxygen in its ground state.

## ■ REFERENCES

Table of Content:

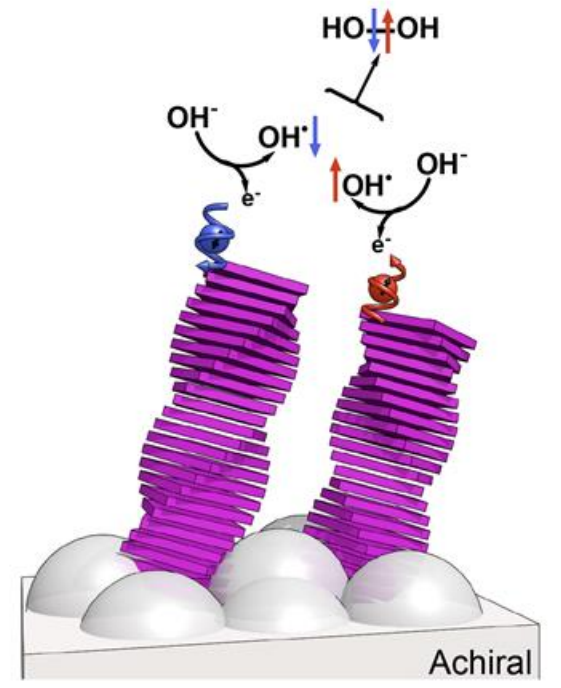

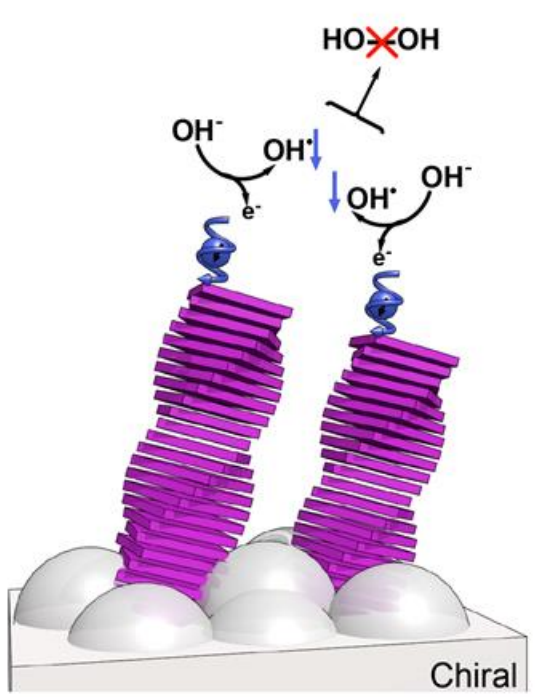